# Classification of Research Citations (CRC)

Bilal Hayat Butt[1], Muhammad Rafi[2], Arsal Jamal[3], Raja Sami Ur Rehman[4], Syed Muhammad Zubair Alam[5] and Muhammad Bilal Alam[6]

[1]*bilal.hayat@nu.edu.pk*, [2]*muhammad.rafi@nu.edu.pk*, [3]*k112348@nu.edu.pk*, [4]*k112352@nu.edu.pk*, [5]*k112092@nu.edu.pk*, [6]*k112351@nu.edu.pk*

National University of Computer and Emerging Sciences NUCES-FAST, Department of Computer Science, Karachi (Pakistan)

**Abstract**

Research is a continuous phenomenon. It is recursive in nature. Every research is based on some earlier research outcome. A general approach in reviewing the literature for a problem is to categorize earlier work for the same problem as positive and negative citations. In this paper, we propose a novel automated technique, which classifies whether an earlier work is cited as sentiment positive or sentiment negative. Our approach first extracted the portion of the cited text from citing paper. Using a sentiment lexicon we classify the citation as positive or negative by picking a window of at most five (5) sentences around the cited place (corpus). We have used Naïve-Bayes Classifier for sentiment analysis. The algorithm is evaluated on a manually annotated and class labelled collection of 150 research papers from the domain of computer science. Our preliminary results show an accuracy of 80%. We assert that our approach can be generalized to classification of scientific research papers in different disciplines.

**Conference Topic**

Citation and Co-Citation Analysis.

**Introduction**

Research is a continuous phenomenon. It is recursive in nature. Every research is based on some earlier research outcome. A general approach in reviewing the literature for a problem is to categorize earlier work for the same problem as positive and negative citations. Tanguy (2009) argues that the multi category classification scheme is needed for Humanities and Social Science (HSS). Author claims that the negative citations are less than 10% in HSS, however, for scientific research, these are not very rare. Scientific research provides quantitative results and thus a critique and rebuttal is viable, as compared to qualitative results in HSS research. With the help of CRC, research scholar can easily classify the type of reading materials, along with other researcher's point of view in citing the same paper. CRC will help the scholar to gather information at a much faster pace than conventional approach of manually looking up and classifying research papers. They can look up for the type (Class) of reading material required namely positive or negative, thereby, minimizing the time to find relevant research for a topic of interest. The goal of CRC is to be used as a resource for assisting researcher scholars.

This paper is organized as follows, it first introduces some formal terms and preliminaries. In Related Work section we argue about various other related researches and discusses similarities and differences with our approach. Then we define our methodology beginning with description of our dataset, extraction techniques adopted, selection of features, and lastly running classification. Further we analyse our experimental results based on which we provide conclusion.

**Preliminaries and Formalism**

In this section we introduce some *formal terms* and *preliminaries* for better understanding of technicalities in paper.

*Terms used with citation analysis*

In *Citation Analysis*, *root paper or cited paper* is the research paper that is being cited in another research paper. *Citation* is a quotation that is used to refer a root paper. *Citing Paper* is the research paper that consists of a reference to the root paper. *Cited Area* is the paragraph in citing paper with citation of root paper. Terms are inspired from Nanba (2000) and explained in Figure 1. *Referencing Style* are the writing styles that one uses to organize the information when writing a research paper.

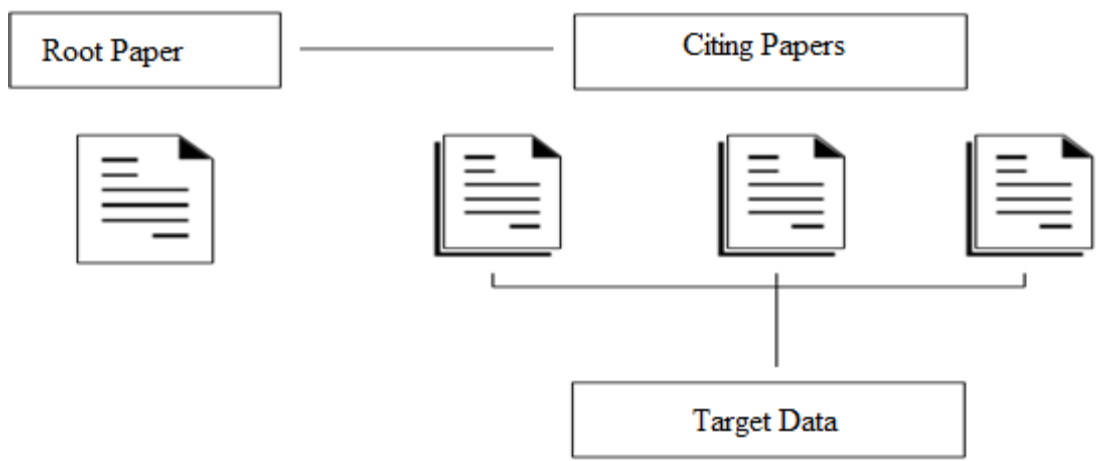

**Figure 1. Citation Analysis**

*Terms used with sentiment analysis*

In *Text Categorization, Tokenization* is the process that breaks down sentences of text into word(s) for further processing. It is done using *Regular Expressions, which* is a generic way of representing a string to be searched from a large corpus of text. *Sentiment Analysis* is done to classify the corpus into positive or negative categories based on the words extracted during tokenization. *Lexicon* is a dictionary that consists of key words which are divided into positive and negative categories. *Positive Incrementer* and *Negative Incrementer* are dictionaries, besides the Lexicon that were used in our approach, consisting of phrases of words that were observed to be inclining a particular sentence to have an overall positive or negative weight.

*Terms used with experimental study*

In our experimental study, *Target Data* is the data extracted from the cited paragraph. *Citation Types are* the different categories into which target data are placed after sentiment analysis. *Naïve-Bayes Classifier* is a popular machine learning algorithm for text categorization. *Monte-Carlo Technique* is a computational algorithm that rely on repeated random sampling. *Precision, Recall and F1-score* are calculated, taken from Manning (2008)**.**

*Precision* is the fraction of retrieved instances that are relevant, calculated through the equation:
$$P = |\{Relevant\ Documents\} \cap \{Retrieved\ Documents\}| / |\{Retrieved\ Documents\}|$$

*Recall* is the fraction of relevant instances that are retrieved, calculated through the equation:
$$R = |\{Relevant\ Documents\} \cap \{Retrieved\ Documents\}| / |\{Relevant\ Documents\}|$$

*F- Measure* is a measure that combines precision and recall. It is the harmonic mean of precision and recall. The balanced F1-score, calculated through the equation:
$$F1 = 2*(Precision*Recall) / (Precision + Recall)$$

**Related Work**

Authors have taken different strategies in approaching citation analysis, Tanguy (2009) provides a detailed overview. One of the approaches focuses on identifying citation categories or classification schemes, while other focuses on identifying cue words that can help better annotation and classification of papers.

*Classification of Citations*

Nanba (2000) classified the cited papers into three categories, manually, using cue words. Author discusses a prototype system called PRESRI that relies on citation relationships. Keeping this research as our base we enhanced and automated the procedure. We applied different methods to classify the papers using lexicon dictionaries instead of cue words.

Tanguy (2009) discusses an automated technique to identify the citation and uses linguistic cues for analysis of French humanities articles through natural language processing techniques. The approach presented is quite similar to CRC in terms of the lexicon used to identify which category a particular sentence belongs to. However, their approach is limited to APA style referencing only, while we have extended it to AMA and IEEE, as well.

Cohen (2006) discusses importance of automated classification of document citations in reducing the time spent by experts in reviewing journal articles. The paper proposes the use of classification of research papers in the field of medicine; specific to the field of drugs and their use in treatment of diseases. Through an automated classification process, the paper puts forward a review system for a selected list of drugs. Thus classifying the drug to be either positive or negative, against a disease. Similar to the approach of CRC, the paper looks for the cited areas which contains the reviews of the drug. The paper then extracts these areas to perform classification. The domain of their research is specific to medical science whereas our approach can be generalized to variety of disciplines.

Mostly techniques reviewed are focused on one discipline while we assert that our approach can be generalized to classification of scientific research papers in different disciplines, since, instead of cue words for a specific domain, we are using generalized sentiment lexica used in Evert (2014). Author identified sentiment lexica as one of the most important features along with bag-of-words unigrams and bigrams, for machine learning classifier.

*Classification Schemes*

Agarwal (2010) developed an eight-category classification scheme, annotated using that scheme, developed and evaluated the supervised machine learning classifiers using annotated data. As discussed in the paper, inter-annotator agreement could not be reached for overlapping categories and author suspect that this issue could increase with huge collection of articles. To overcome this issue, we have combined the overlapping categories.

Teufel (2006) proposed a scheme of 12 categories for any citation being made. Table 1 shows another such scheme suggested by Spiegel (1977) with 13 different motivations which could lead an author to cite any research paper. This scheme is discussed in detail, because it is based on scholarly articles published in science studies.

**Table 1. Reasons for Citing a Paper**

| |
|---|
| 1. Cited Paper provides historical facts regarding undergoing Research Question. |
| 2. Continuing a Research from point where Cited paper finished. |
| 3. Citing paper to use its ideas, definitions, terms in a Research |
| 4. Citing a paper to refer to data also used in Current Research. |
| 5. Citing a paper to refer to data it used and to draw similarities from the Data used. |
| 6. Citing Paper contains Data and Material used throughout different phases |
| 7. Citing Paper to adopt part/full methodology it adopted for a certain task. |
| 8. Citing paper verified/proved a statement or enlightens with its details. |
| 9. Citing Paper evaluated positively. |
| 10. Citing Paper evaluated negatively. |
| 11. Ongoing Research giving proof of statement in Cited Paper. |
| 12. Ongoing Research giving rebuttal of statement in Cited Paper. |
| 13. Giving a new interpretation to the findings/statements in Cited Paper. |

In our approach we have grouped these citation types into three generalized sentiment types:
- TYPE-I:   Positive
- TYPE-II:  Negative
- TYPE-III: Neutral

Referring to Table 1, Type-I strictly refers to 1, 2, 3, 4, 5, 6, 7, 8, 9, 11, Type-II strictly refers to 10, 12, 13 and Type-III refers to 4. However 1, 6 may fall under all three Types and 2, 3 may fall under Type-I and Type-II. Such overlapping nature of categories creates dis-agreement between annotators Agarwal (2010) and Tanguy (2009), so we narrowed down the categories.

**Methodology**

In this paper, we propose a novel automated technique, which classifies whether an earlier work is cited as sentiment positive or sentiment negative. Our approach first extracted the portion of the cited text from citing paper. Using a sentiment lexicon we classify the citation as positive or negative by picking a window of at most five (5) sentences around the cited place (corpus). The algorithm is evaluated on a manually annotated and class labelled collection of 150 research papers from the domain of computer science. Complete process is explained in Figure 2, and reference to the example papers are provided. Root paper followed by citing papers.

*Dataset Collection*

We collected a data set consisting of 150 research papers, manually downloaded in pdf format from Google Scholar. To further strengthen our basis for citation types we devised manual annotation of citation corpuses from all research paper in our data set and then manually classifying them into Type-I and Type-II. The results are shared in Table 2.

**Table 2. Distribution of Papers According to Their Types.**

| | Type-I | Type-II | Total |
|---|---|---|---|
| Papers Dataset | 109 | 41 | 150 |

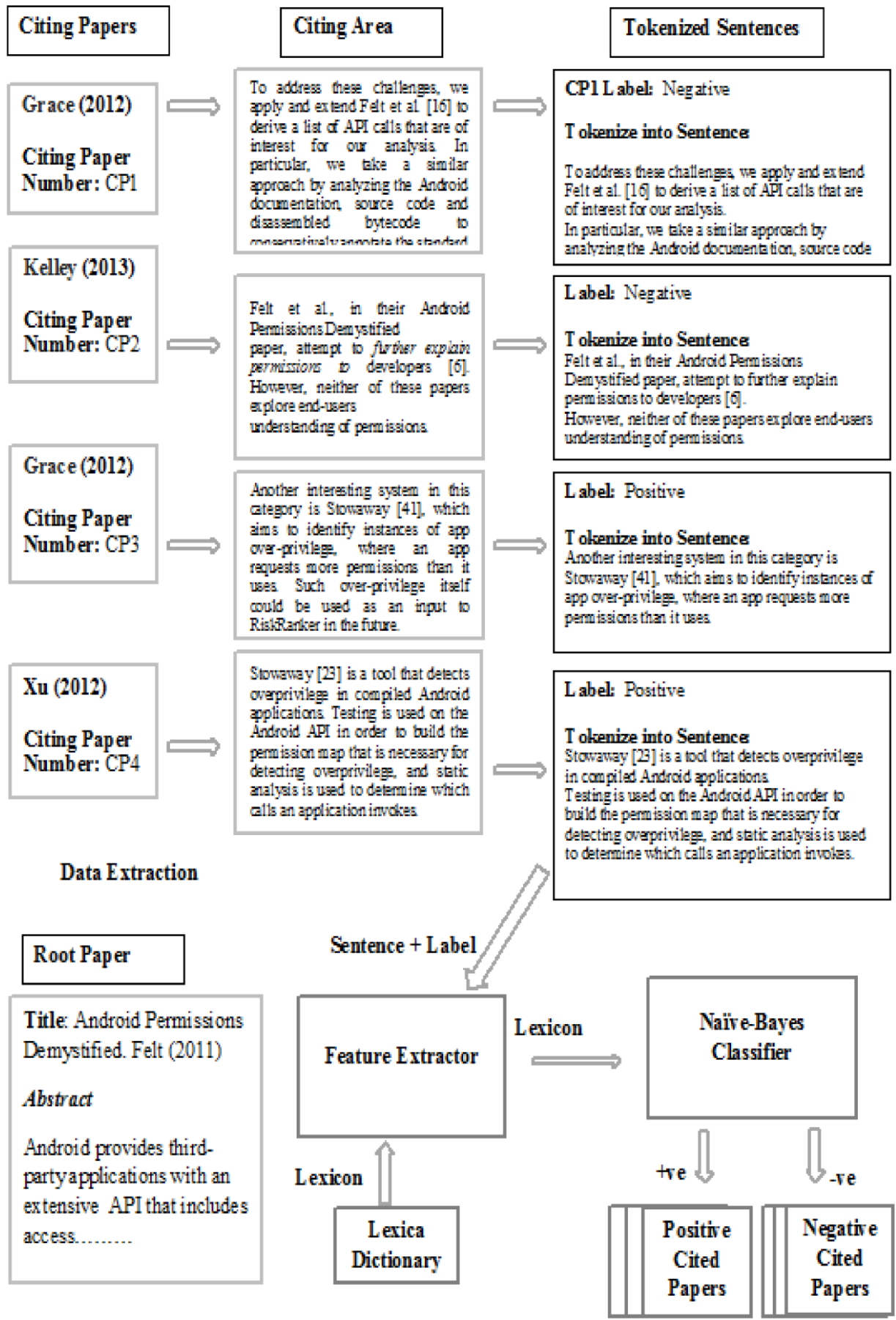

**Figure 2. Overview of CRC**

*Annotation*

All the annotation were done manually by the co-authors and the results were further cross-checked by other co-authors. Inter-annotator agreement was fair, however, to further strengthen our annotation results, we feel that annotation should be done by individuals who have at least a post graduate degree in their fields related to the corpus. In future, we plan to work with PhD Scholars in the field of computer science to provide annotated data in their field of research.

*Referencing Styles*

For every root paper that we queried, we stored its top 10 citing papers. To make our approach more robust we included research papers of different writing styles such as Institute of Electrical and Electronics Engineers (IEEE) Standard Style, American Psychological Association (APA) Style and American Medical Association (AMA) Manual of Style. We summarize variations in referencing style in Table 3, along with percentage of sample papers in our dataset.

**Table 3. Types of Research Paper Collected.**

|  | IEEE | APA | AMA |
| --- | --- | --- | --- |
| Reference Style | [1] Name Paper, Journal. Year | Name (Year). Paper, Journal | 1. Name Paper, Journal. Year |
| In-Text Citation | [1], [1,2] | Name (year) | [1], [1,2] or $_{1-2}$ |
| Percentage of papers | 67% | 15% | 18% |

*Data Conversion Module*

Root paper and all its citing papers were placed in separate folders. The first step was to convert all pdf files in dataset into text format. Python Library 'PDF Miner' Shinyama (2010) was used for file conversion. All converted text files were placed in their respective folders.

*Data Extraction Module*

To process data for analysis phase we extract useful information from all text files in the dataset. From root paper we extract the title and the abstract section. Similarly, from citing paper we extract the title of citing paper, reference (number) which points to in-text citations made to root paper, reference to root paper, corpus, continuing reference i.e. reference to any research paper other than root paper in corpus region, an annotated label assigned to each citing paper. First the title of each root paper is extracted, but pdf-to-text conversion tend to modify formatting style, making it difficult to construct a regular expression to identify title of each root paper. Instead we programmed a Scrapping Module API in Python Venthur (2014). Scrapper searches the Input on Google Scholar and successfully scraps the title of target paper. Extracted title and its path on our system directory are stored in a CSV file. We use these root paper titles to search for reference in their respective citing paper. First we extract the reference section in each citing paper, let's call this *cropReference*. Next we search for root paper's title in *cropReference*. Depending on the format style of citing paper a particular regular expression execute to correctly extract reference. Following this, with the help of reference we form another regular expression to extract corpus region of maximum five (5) sentences. We search and remove any sentence in our corpus that includes reference to any paper other than our root paper. All this information is stored in a XML file. Process is explained in Figure 3.

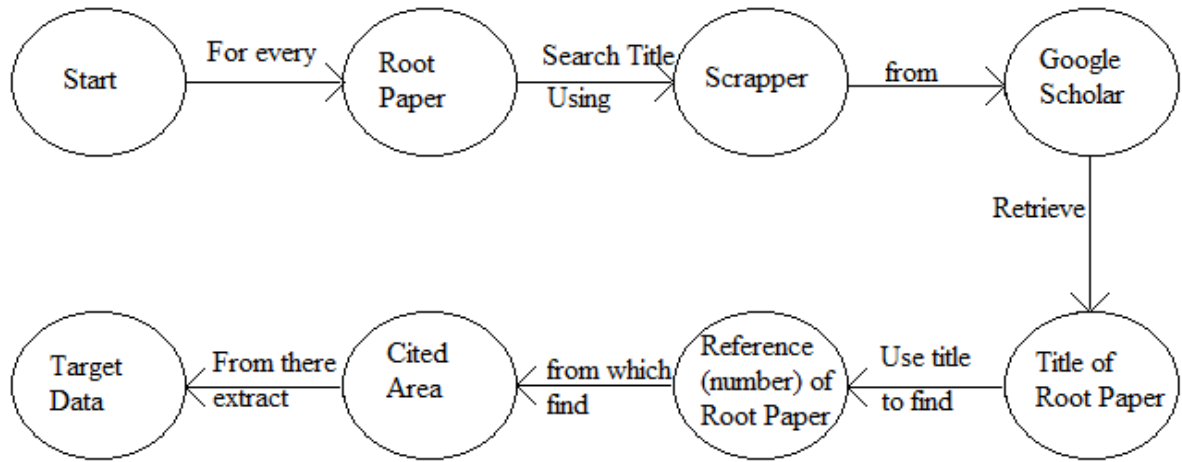

**Figure 3. Overview of Data Extraction Process**

*Parsing XML file*

XML Parser creates a separate file for each citing paper containing its Class (Label), Reference (number) and Corpus. These files were then placed in respective subfolder (Type I, Type II).

*Feature Extractor*

Once the dataset is ready, we made features upon which citing papers will be automatically classified. For this purpose we included following lexicon dictionaries:
- Positive and Negative lexica: Derived from Evert (2014) consisting of approximately 28,000 distinct positive and 31,000 distinct negative words.
- Positive Incrementer and Negative Incrementer: Derived from Nanba (2000). Each includes approx. 75 entries that are unigrams, bigrams or trigram.

Feature Extractor iteratively reads each files. It uses Text Blob library Loria (2014) in Python for tokenization. First corpus is tokenized into sentences to compare with Positive and Negative Incrementer, and then each sentence is tokenized into words to compare with Positive and Negative Dictionaries. For each match and mismatch between token and dictionary we create a feature-set list, composed of Token ('1' for match and '0' for mismatch) and Label of that citing paper.

*Running the Classifier*

We have used Naïve-Bayes Classifier for sentiment analysis, using Scikit-learn library Pedregosa (2011) and NLTK library Bird (2009). Due to limited data, we had to decide what portion of feature-set will go into training and what portion of feature-set will this classifier test. In our first approach we trained and tested our classifier on all of feature-set. Next, we used different techniques to distribute feature-set in training and testing sets.

**Experimental Study**

In this section we discuss our findings for data extraction and classification. For data extraction our results show 86% accuracy, which is not only higher than the estimated precision score of 80% by Tanguy (2009) for syntactic parsing, but also we have generalized the approach for different referencing styles. Few papers from which we could not accurately extract data were either due to pdf-to-text conversion limitations for images and remaining were because referencing style were not correctly followed by the author. Results are explained in Table 4.

**Table 4. Accuracy of Data Extraction.**

|  | Successfully extracted | Issue with pdf-to-text conversion | Referencing Style not followed | Total count |
|---|---|---|---|---|
| Papers Dataset | 130 | 5 | 15 | 150 |

After extracting the data from the citing papers, the data was analysed with the help of Naïve-Bayes classifier. Our preliminary results show an accuracy of 80%, explained in Table 5.

**Table 5. Results Obtained Using Naïve-Bayes Classifier.**

| Category | Precision | Recall | F1-score | Support |
|---|---|---|---|---|
| Type I | 0.84 | 0.94 | 0.89 | 109 |
| Type II | 0.25 | 0.10 | 0.14 | 21 |
| Average/Total | 0.75 | 0.81 | 0.77 | 130 |

The reason for low recall for Type II was because we had only 21 papers that were manually annotated to be negative while the rest of the 109 papers were annotated to be positive. Due to lack of data our classifier couldn't accurately train itself on Type II papers. To further strengthen our result and remove low sampling bias of Type II papers we have used two approaches. In our first approach, the data set was adjusted to 42 research papers. For testing and training we used 50% data. To balance Type I and Type II papers, the technique was applied on different window sizes of Type I papers, while using all 21 papers of Type II. The windows were set to 0-21, 22-43, 44-65, 66-87, 88-109 and 110-125. Different accuracy were obtained at each window and the average accuracy was approximately 78%. Results are explained in Table 6.

**Table 6. Result and Distribution in Feature-Set.**

| No. | Test Type | Window (distribution of Type I papers) | Accuracy |
|---|---|---|---|
| 1 | Train=21, Test=21 | 0-21 | 81% |
| 2 | Train=21, Test=21 | 22-43 | 77% |
| 3 | Train=21, Test=21 | 44-65 | 71% |
| 4 | Train=21, Test=21 | 66-87 | 85% |
| 5 | Train=21, Test=21 | 88-109 | 67% |
| 6 | Train=21, Test=21 | 110-125 | 85% |
|  |  | Average | (78%) |

Secondly, the data was analysed using the Monte-Carlo Technique, to obtain stability in classification. Further changes were made to the Type I windows accordingly, explained in

Table 7. The window size was now set to 0-21, 1-22, 2-23 and so on, while keeping all 21 papers of Type II. The average accuracy obtained was 72%.

**Table 7. Result with random sampling.**

| No. | Test Type | Window (distribution of Type I papers) | Accuracy |
|---|---|---|---|
| 7 | Train=21, Test=21 | 0-21, 1-22, 2-23, 3-24, 4-25, 5-26, … , 22-43, 23-44, 24-45, 25-46, 26-47, … , 35-56, 36-57, 37-58, 38-59, 39-60, … , 50-71, 51-72, 52-73, 53-74, 54-75, … , 60-81, 61-82, 62-83, 63-84, 64-85 … , 75-96, 76-97, 77-98, 78-99, 79-100, … , 85-106, 86-107, 87-108, 88-109 … , 96-117, 97-118, 98-119, 99-120, … | Average (72%) |

**Conclusion and Future Work**

The purpose of our paper was to help the research scholars by minimizing the time required to find the relevant research, for a topic of interest. In CRC, we proposed to classify the citations into three categories i.e. Type-I (Positive), Type-II (Negative) and Type-III (Neutral). A researcher now if desires to know any further advancements in paper under study can directly refer to Type-I papers, or if he wishes to know any research giving rebuttal of findings, in paper under study, can refer to Type-II paper. Our preliminary results show an accuracy of 80%. We assert that the technique can be generalized to classification of scientific research papers. Currently, support for Type-III papers in CRC is in progress, and we are working on a lexicon dictionary with neutral words. In future, we plan to provide a web portal to assist the research scholars in automatically searching and downloading citing papers, for a root paper, and classification of citing papers into sentiment categories.

**References**

Agarwal, S., Choubey, L., & Yu, H. (2010). Automatically classifying the role of citations in biomedical articles. *In AMIA Annual Symposium Proceedings (Vol. 2010, p. 11). American Medical Informatics Association.*

Bird, S., Klein, E., & Loper, E. (2009). Natural language processing with Python. *O'Reilly Media, Inc*.

Cohen, A. M., Hersh, W. R., Peterson, K., & Yen, P. Y. (2006). Reducing workload in systematic review preparation using automated citation classification. *Journal of the American Medical Informatics Association, 13(2), 206-219.*

Evert, S., Proisl, T., Greiner, P., & Kabashi, B. (2014). SentiKLUE: Updating a Polarity Classifier in 48 Hours. *SemEval, 551.*

Loria, S. (2014), *Textblob: Simplified Text Processing.* Retrieved 4 May 2015 from: http://textblob.readthedocs.org/en/dev/

Manning, C. D., Raghavan, P., & Schütze, H. (2008). Introduction to Information Retrieval *(Vol. 1, p. 496). Cambridge: Cambridge university press.*

Nanba, H., Kando, N., & Okumura, M. (2000). Classification of research papers using citation links and citation types: Towards automatic review article generation. *11th ASIS SIG/CR Classification Research Workshop*

Pedregosa, F., Varoquaux, G., Gramfort, A., Michel, V., Thirion, B., Grisel, O, & Duchesnay, E. (2011). Scikit-learn: Machine learning in Python. *The Journal of Machine Learning Research, 12, 2825-2830.*

Shinyama, Y. (2010) *PDFMiner: Python PDF parser and analyzer*. Retrieved on 11 June 2015 from: http://www.unixuser.org/~euske/python/pdfminer/

Spiegel-Rösing, I. (1977). Science studies: Bibliometric and content analysis. *Social Studies of Science, 97-113*.

Tanguy, L., Lalleman, F., François, C., Muller, P., & Séguéla, P. (2009). RHECITAS: citation analysis of French humanities articles. *In Corpus Linguistics 2009 (pp. http-ucrel)*.

Teufel, S., Siddharthan, A., & Tidhar, D. (2006, July). Automatic classification of citation function. *In Proceedings of the Conference on Empirical Methods in Natural Language Processing (pp. 103-110). Association for Computational Linguistics.*

Venthur B. (2014) GScholar: Query Google Scholar with Python. Retrieved 5 April 2015 from: http://github.com/venthur/gscholar/

*Reference for papers used in explanation (Figure 2).*

Felt, A. P., Chin, E., Hanna, S., Song, D., & Wagner, D. (2011, October). Android permissions demystified. In Proceedings of the 18th ACM conference on Computer and communications security (pp. 627-638). ACM.

Grace, M. C., Zhou, W., Jiang, X., & Sadeghi, A. R. (2012, April). Unsafe exposure analysis of mobile in-app advertisements. In Proceedings of the fifth ACM conference on Security and Privacy in Wireless and Mobile Networks (pp. 101-112). ACM.

Grace, M., Zhou, Y., Zhang, Q., Zou, S., & Jiang, X. (2012, June). Riskranker: scalable and accurate zero-day android malware detection. In Proceedings of the 10th international conference on Mobile systems, applications, and services (pp. 281-294). ACM.

Kelley, P. G., Cranor, L. F., & Sadeh, N. (2013, April). Privacy as part of the app decision-making process. In Proceedings of the SIGCHI Conference on Human Factors in Computing Systems (pp. 3393-3402). ACM.

Xu, R., Saïdi, H., & Anderson, R. (2012, August). Aurasium: Practical Policy Enforcement for Android Applications. In USENIX Security Symposium (pp. 539-552).